\documentstyle[sprocl]{article}

\def\Journal#1#2#3#4{{#1} {\bf #2}, #3 (#4)}

\def\PLB{{\em Phys. Lett.}  B}
\def\PRL{\em Phys. Rev. Lett.}

\def\be{\begin{equation}}
\def\ee{\end{equation}}
\def\bea{\begin{eqnarray}}
\def\eea{\end{eqnarray}}

\begin{document}

\title{LORENTZ SYMMETRY VIOLATION AT PLANCK SCALE,
COSMOLOGY AND SUPERLUMINAL PARTICLES}

\author{L. GONZALEZ-MESTRES}

\address{L. P. C. Coll\`ege de France,
11 pl. Marcellin-Berthelot\\ 75231 PARIS Cedex 05 , FRANCE\\
Laboratoire d'Annecy-le-Vieux de Physique des Particules, B.P. 110\\
74941 ANNECY-LE-VIEUX Cedex, FRANCE}


\maketitle\abstracts{Although Lorentz symmetry has been tested at low energy
with extremely good accuracy, its validity at very high energy is much less
well established. If Lorentz symmetry violation (LSV) is energy-dependent
(e.g. $\propto ~E^2$), it can be of order 1 at Planck scale and 
undetectable at GeV scale or below. 
Similarly, superluminal particles with positive mass and energy 
({\bf superbradyons}) can exist and be the ultimate building blocks of matter. 
We discuss a
few cosmological  
consequences of such a scenario, as well as possible experimental tests.} 

\section{Lorentz symmetry violation and superluminal 
particles}

~~~~~{\it "Experiment has provided numerous facts justifying the following
generalization: absolute motion of matter, or, to be more precise, the
relative motion of weighable matter and ether, cannot be disclosed. All that
can be done is to reveal the motion of weighable matter with respect to
weighable matter" (H. Poincar\'e, 1895)}

{\it "Such a strange property seems to be a real coup de pouce presented by
Nature itself, for avoiding the disclosure of absolute motion...
I consider quite probable that optical phenomena depend only on
the relative motion of the material bodies present, of the sources of
light and optical instruments, and this dependence is not accurate...
but rigorous.
This principle will be confirmed with increasing precision,
as measurements
become more and more accurate" (H. Poincar\'e, 1901)}

{\it "The interpretation of geometry advocated here cannot be
directly applied
to submolecular spaces... it might turn out that such an extrapolation is
just as incorrect as an extension of the concept of temperature to particles
of a solid of molecular dimensions" (A. Einstein, 1921)} 
 
\subsection{Status of the Poincar\'e relativity principle}

The Poincar\'e relativity principle \cite{Poincare95,Poincare01} has been
confirmed by very accurate low-energy tests \cite{Lam,Hills}, but its
validity at much higher energies is not obvious \cite{Gon1,Gon2}. 
The possibility that special relativity could fail at small distance
scales was already considered by A. Einstein \cite{Ein21}: it is remarkable
that the relativity principle 
holds at the energies attained by particle accelerators.
Experiments devoted to the highest-energy cosmic rays may provide
crucial tests of Lorentz symmetry \cite{Gon1}.
 
\subsection{Lorentz symmetry violation (LSV)}

Lorentz symmetry can be broken introducing a local absolute rest frame
(the vacuum rest frame, VRF)
and a fundamental distance scale $a$ 
\cite{Gon1}. If LSV follows a $\propto ~E^2$ law ($E$ = energy), it
can be $\approx 1$ at Planck scale and $\approx ~10^{-40}$ at 
the $\approx ~100~MeV$ scale, escaping all low-energy
tests of Lorentz symmetry. But a $\approx ~10^{-6}$
LSV at Planck scale can produce \cite{Gon1} obervable effects
at the highest
cosmic-ray energies ($\approx ~10^{20}~eV$). 
If $k$ is the wave vector,
nonlocal models lead in the VRF \cite{Gon1}
to
a deformed relativistic kinematics which for $k~a~\ll ~1$ gives:
\equation
E ~~ \simeq ~~ c~(p^2~+~m^2~c^2)^{1/2}
~-~(c~\alpha /2)~(p~k~a)^2~(p^2~+~m^2~c^2)^{-1/2}
\endequation
where $p$ stands for momentum, $m$ for mass and $\alpha $ is a 
positive constant.

\subsection{Deformed relativistic kinematics (DRK)}

Contrary to the $TH\epsilon \mu $ model \cite{Will}, DRK preserves relativity 
in the limit $k~\rightarrow 0$ .
A fundamental question is that of the universality of 
$\alpha $ : is $\alpha $ the same for all bodies, or does it depend on 
the object under consideration? If $c$ is universal and
$\alpha ~\propto ~m^{-2}$ , equation (1)
amounts to a relation between $E/m$ and $p/m$ , as in 
relativistic kinematics. From a naive soliton model \cite{Gon3}, 
we inferred that: a) $c$ is
expected to be universal up to very small corrections ($\sim ~10^{-40}$)
escaping existing bounds; b)
an
approximate rule can be to take $\alpha $ universal for leptons, gauge bosons
and light hadrons (pions, nucleons...) and assume a $\alpha \propto m^{-2}$
law for nuclei and heavier objects, the nucleon mass setting the scale.

\subsection{Cosmic superluminal particles (CSL)}

If Lorentz symmetry is broken at Planck scale, nothing prevents the existence 
of particles with positive mass and energy and critical speed in vacuum $c_i$
(the subscript $i$ stands for the $i$-th superluminal sector) 
much larger than the speed of light $c$ \cite{Gon2}. Such particles 
({\bf superbradyons}) could be the ultimate building blocks of matter from
which, for instance, strings would be made. They can satisfy the same 
kinematics as "ordinary" particles, but replacing the speed of light $c$ by
the new critical speed $c_i$ , and interact weakly with "ordinary" matter.
Nonlocal models at Planck scale
may be the limit of an underlying superluminal
dynamics in the limit $c~c_i^{-1}~\rightarrow ~0$ .
CSL can possibly propagate in vacuum just like photons in a perfectly
transparent crystal. 

\section{Some cosmological implications}

It was suggested \cite {Bacry}, using a different DRK from (1), 
that DRK could explain the dark matter problem: 
the non additivity of rest energy for non interacting systems at rest would
account for the illusion of a missing mass. But it was later argued 
\cite{Fern} that the
effect would actually be opposite to observation. However, both 
authors use a model where the additive quantity,
instead of energy, is:
\equation
F~(m~,~E )~~=~~2~\kappa ~(m)~~sinh~[2^{-1}~\kappa ^{-1}~(m)~E ]
\endequation
and the constant $\kappa $ (similar to the parameter $\alpha $ of our model)
has a universal value. There is no fundamental 
reason for this universality and similar arguments to those developed for
our DRK model would suggest \cite{Gon3} $\kappa ~\propto ~m$ , restoring 
the additivity of rest energy for large non interacting systems at rest.

A generalization of Friedmann equations in the presence of superluminal
sectors of matter can be built \cite{Gon4} 
and does not present inconsistency with data. Superluminal particles may
actually be most of the cosmic (dark) matter.

\end{document}